\documentclass[journal]{IEEEtran}
\ifCLASSINFOpdf
\else
\fi
\usepackage[scriptsize]{caption}
\usepackage{graphicx, dblfloatfix}
\usepackage{epstopdf}
\usepackage{amsmath}
\usepackage{amssymb}
\usepackage{graphics}
\usepackage{cite}
\usepackage{amsfonts}
\usepackage{textcomp}
\usepackage{multicol}
\usepackage{multirow}
\usepackage{wrapfig}
\usepackage{color}
\usepackage{fixltx2e}
\usepackage{subfig}
\begin{document}
\title{A Polarization Multiplexed Carrier based Coherent Link with Adaptive Polarization Control}
\author{Rashmi Kamran, Sana Naaz, Sarath Manikandan, Sandeep Goyal, Rakesh Ashok, and Shalabh Gupta\thanks{Authors are with Department of Electrical Engineering, IIT Bombay, Powai, Mumbai 400076, India e-mail: (rashmikamran@ee.iitb.ac.in).}}
\maketitle
\begin{abstract}
Transmitting polarization multiplexed carrier makes the receiver of a coherent system local oscillator-less and frequency offset-free. A polarization multiplexed carrier based self-homodyne (PMC-SH) system with an adaptive polarization control (PC) can replace pulse amplitude modulation (PAM-4) data center interconnects. An adaptive PC technique is practically implemented by using an electrically controlled PC along-with control circuitry for PMC-SH systems. The de-multiplexing of the carrier and the modulated signal by using this technique is validated through simulations for a 50\,Gbaud PMC-SH\, quadrature phase shift keying (QPSK) system with 20\,km standard single mode fiber (SSMF). We successfully demonstrate 16\,Gbaud PMC-SH systems with adaptive PC for 10\,km SSMF channel. A bit error rate (BER) of 5.9 $\times$ 10$^{-5}$ is achieved with 32\,Gb/s PMC-SH\,QPSK system without any signal processing while a BER of 8.7$\times$ 10$^{-3}$ is achieved with a 64\,Gb/s PMC-SH\,quadrature amplitude modulation (16\,QAM) system after equalization.
\end{abstract}

% Note that keywords are not normally used for peerreview papers.
\IEEEpeerreviewmaketitle
\section{Introduction}
\IEEEPARstart{T}{here} is an urgent need to develop low power high-speed short-reach interconnects for datacenter applications. \mbox{PAM-4} is presently being used to achieve 100\,Gbps rates per wavelength for such requirements \cite{pam4}. To improve data rates further, low power coherent techniques are being considered \cite{firstofc2,newanalog}. In general, power consumption requirements of receivers for coherent links make them  prohibitive for short-reach applications. Among other things, typically a coherent link receiver requires a local-oscillator (LO) laser and significant additional signal processing overheads to correct for frequency offsets. In addition, both the transmit and the receive lasers need to be of high quality due to narrow linewidth requirements. 

Polarization multiplexing of the carrier at the transmitter of a coherent link offers several advantages that lead to significant power savings, as (i) it obviates the need for an LO laser at the receiver; (ii) signal processing requirements are significantly reduced in the absence of frequency offsets between the modulated signal and the LO; and (iii) the system can tolerate broad-linewidth lasers as the relative phase offset between the signal and the modulated carrier is independent of the carrier phase noise in such a system\cite{WHYSH4,RPPMC11}. In comparison of an intensity-modulation direct-detection (IMDD) system, such as a PAM-4 system, a polarization multiplexed carrier based coherent self-homodyne (PMC-SH) system, such as a 16-QAM system, offers double the data-rate available for the given baud-rate with similar SNR and bandwidth requirements for the electronics. In addition, due to coherent nature, dispersion effects can be treated as linear impairments that makes signal processing easier at high baud-rates (and dispersion effects can also be mitigated). 

Although the SH systems can be implemented in many ways \cite{RPPMC12,RPPMC13}, most implementations have some practical limitations. The PMC-SH systems appear to be most promising (for example, a 16-QAM 50\,Gbaud system can be used for 200\,Gbps data rates). However, such SH systems had also been considered as impractical until now because of the requirement of separation of the signal from the polarization multiplexed carrier for coherent demodulation. In this work, we present an adaptive polarization control technique for de-multiplexing of the carrier and the modulated signal at the receiver for reception of QPSK and 16-QAM signals. The proposed approach uses low-frequency electronics for polarization control.  With this, for the first time we have demonstrated a practical PMC-SH system for high-capacity low-power datacenter interconnects.

\section{Polarization diversity based self-homodyne system with adaptive polarization control}
\begin{figure*}[htbp]
	\centering
	\hspace{-0.6cm}
	\includegraphics[scale=0.57]{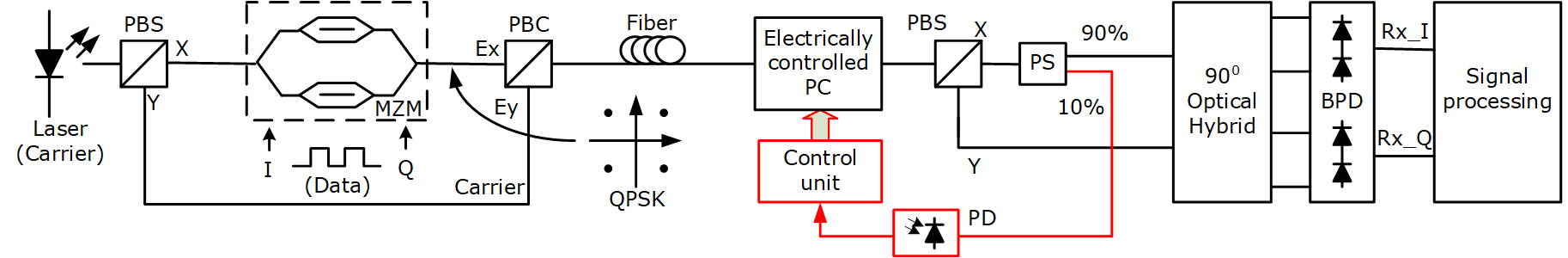}
	%\vspace{-0.2cm}
	\caption{ An PMC-SH\,QPSK/16QAM and adaptive polarization control. PBS/PBC: Polarization Beam Splitter/Combiner; MZM: Mach Zehnder Modulator; PD: Photo Detector; PC: Polarization Controller; PS: Power Splitter; and BPD: Balanced Photo Detector.}
	\label{block}
	%\vspace{-0.8cm}
\end{figure*}

In a polarization diversity based SH system, the carrier is polarization multiplexed along-with the modulated signal. As shown in Fig.\,\ref{block}, the laser output is split into two orthogonal polarizations using a polarization beam splitter (PBS) at the transmitter. One of the outputs is (QPSK or 16\,QAM) modulated using a Mach Zehnder IQ modulator. The modulated signal is significantly weaker than the unmodulated carrier from the other polarization (due of insertion loss of the modulator). The modulated (signal) and the unmodulated (carrier) components are combined using a polarization beam combiner\,(PBC) and sent to the optical channel. At the receiver end, after polarization control (PC), a PBS splits the received signal into two polarizations (ideally separates the carrier and the modulated signals perfectly). The optical hybrid and the balanced photodetectors mix the two outputs to receive the downconverted electrical signals. Received I and Q signals are processed for the compensation of channel and system impairments.
\par In a practical scenario, the received signal consists of a combination of the modulated signal and the carrier due to polarization mode dispersion and polarization rotation. The state of polarization at the input of the PBS fluctuates rapidly due to time-varying temperature gradients and other such effects. Real-time adaptive polarization control is required before the PBS to separate the two perfectly.  Before the PBS, maximization of power in one polarization and/or minimization of power in the other polarization can result in the desired state of polarization (as the carrier power in one polarization is significantly higher than the modulated signal power in the orthogonal polarization). This approach works well specifically for short distance links, in which the effects of PMD (polarization mode dispersion) are insignificant. 

\par An electrically controlled endless polarization controller (EPC) can be used with a control unit to achieve the desired state of polarization by maintaining minimum power in one polarization\cite{pol3}. A fraction (approximately 10\%) of the power in one of the polarizations from the PBS is coupled to a photodetector. The photodetector output is provided to the control unit. Gradient descent algorithm is used by the controller for minimizing the photodetector output by adjusting the control voltages going to the EPC.  
\section{Results}
\subsection{Simulation results}
This scheme was first validated through simulations performed in VPItransmissionMaker$^{TM}$ (adaptive control implemented by using script editor feature) for 50\,Gbaud PMC-SH\,QPSK system. Block diagram shown in Fig.\ref{block} was modeled with laser power of 10\,mW and modulator insertion loss of 12\,dB. Dispersion coefficient of 16\,ps/km.nm, dispersion slope of 0.08$\times$10$^{-3}$s/m$^3$ and attenuation of 0.2\,dB/km were considered for the fiber channel. Polarization mode dispersion and non-linearity were "ON" for simulations. Optical signal to noise power (OSNR) was at 25\,dB which was obtained by adding an OSNR module in the channel. The shot noise and thermal noise were also "ON" for simulations.
\begin{figure}[!h]
	\centering
	%\vspace{-0.3cm}
		\includegraphics[width=\columnwidth]{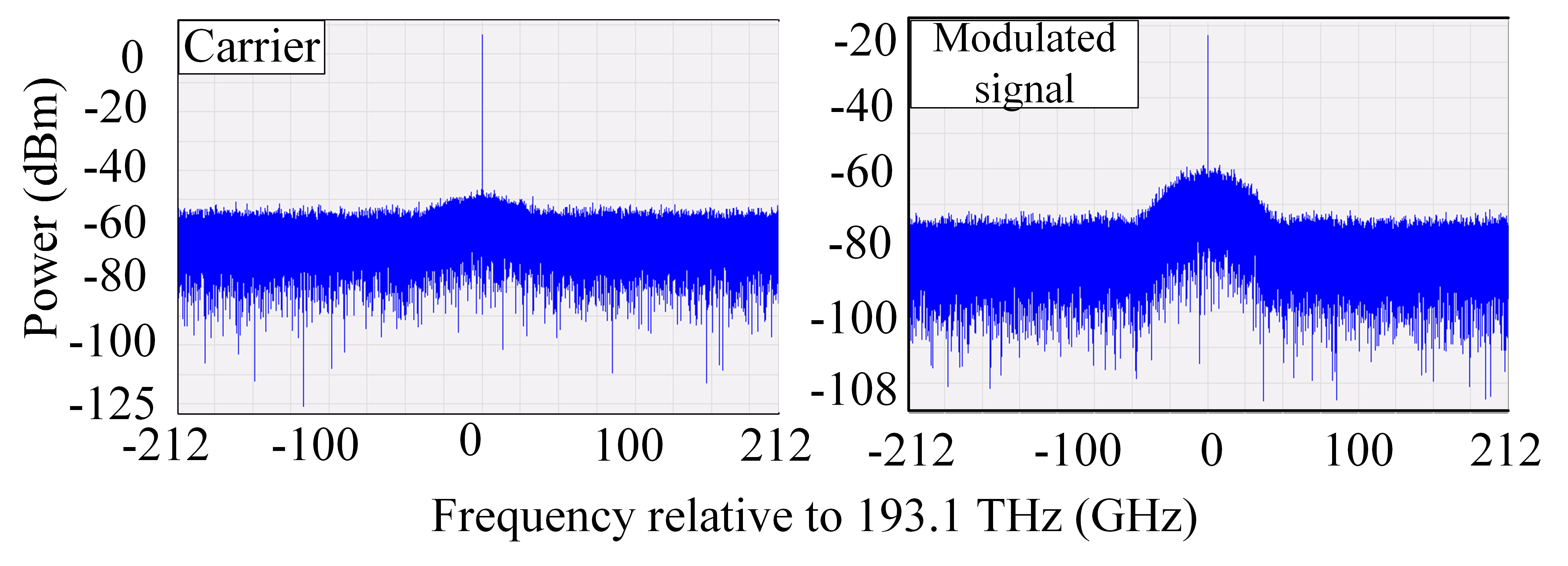} \\
	\caption{Spectrum of X and Y polarizations for an PMC-SH\,QPSK system without minimization of power in one polarization for date rate: 50\,Gbaud, distance: 20\,km and OSNR: 25\,dB. The carrier is mixed with the modulated signal and the modulated signal is mixed with the carrier.}
	\label{pol7}
\end{figure}
\begin{figure}[!h]
		\centering
	\includegraphics[width=\columnwidth]{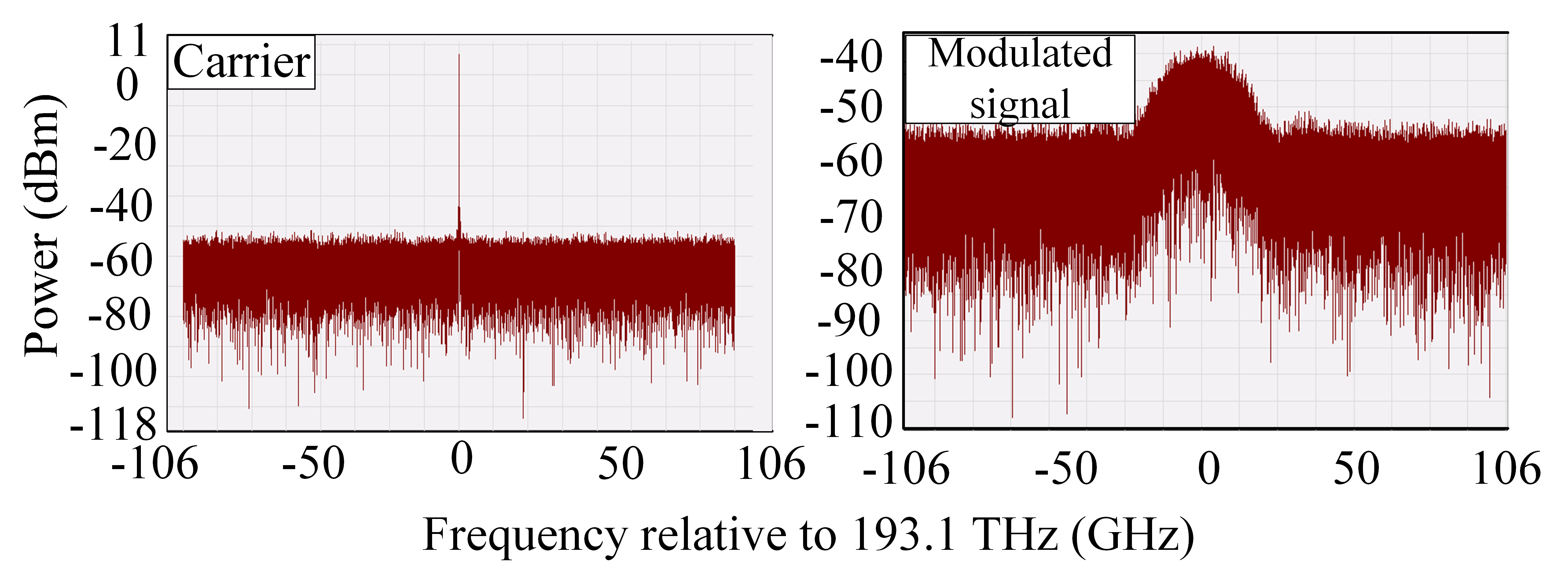}
	\caption{Spectrum of X and Y polarizations for an PMC-SH\,QPSK system with minimization of power in one polarization for date rate: 50\,Gbaud, distance: 20\,km and OSNR: 25\,dB. Separated carrier and separated modulated are presented.}
	\label{pol7no}
\end{figure}
\par As illustrated by the simulation results in Figs.\,\ref{pol7} and \ref{pol7no}, the carrier and the modulated signal are separated by minimizing the power in one of the polarizations. This adaptive controlling is successfully validated through simulations in the presence of significant dispersion and for high data rates (50\,Gbaud SH-QPSK system). 
%\begin{figure}[!h]
	%\vspace{-2.5cm}
%	\centering
%		\includegraphics[width=0.7\columnwidth]{Figures/50gwithpclo.png} 
%	\caption{Spectrum of X polarization for an SH-QPSK system with  minimization of power in one polarization for date rate: 50\,Gbaud, distance: 20\,km and OSNR: 25\,dB. Separated carrier after minimization.}
%	\label{pol7}
	%\vspace{-0.4cm}
%\end{figure}
%\begin{figure}[!h]
	%\vspace{-2.5cm}
%	\centering
%\includegraphics[width=0.7\columnwidth]{Figures/50gwithpcsignal.png}
	%\caption{Spectrum of Y polarization for an SH-QPSK system with  minimization of power in one polarization for date rate: 50\,Gbaud, distance: 20\,km and OSNR: 25\,dB. Separated  modulated signal after minimization.}
	%\label{pol7}
	%\vspace{-0.4cm}
%\end{figure}
\subsection{Experimental results}
Experiments have been performed for 10\,Gbaud and 16\,Gbaud PMC-SH systems with 10\,km standard single mode fiber as the optical channel. The baud-rate and  fiber length have been chosen keeping the bandwidth limitations and mismatches in the electrical components (particularly the RF cables) in mind.
\begin{figure*}[htbp]
	\centering
	\includegraphics[scale=0.7]{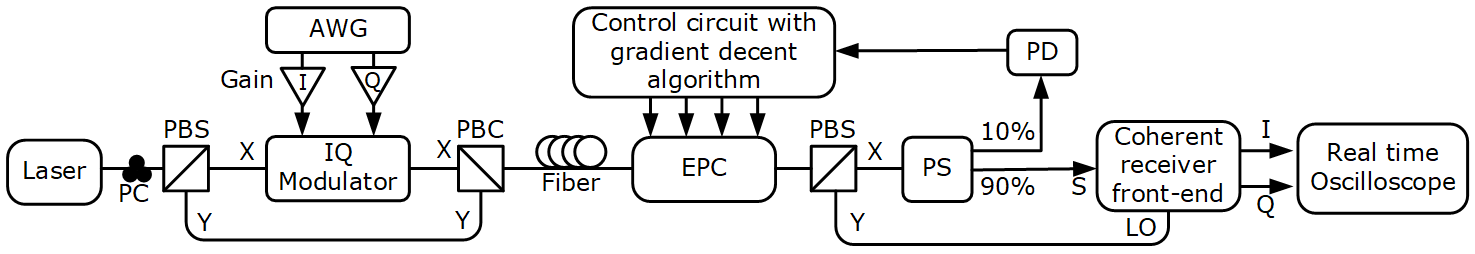}
	%\vspace{-0.2cm}
	\caption{ An PMC-SH\,QPSK/16QAM with adaptive polarization control. PBS/PBC: Polarization Beam Splitter/Combiner; MZM: Mach Zehnder Modulator; PD: Photo Detector; PC: Polarization Controller; PS: Power Splitter; and BPD: Balanced Photo Detector.}
	\label{block1}
	%\vspace{-0.8cm}
\end{figure*}
Experimental setup block level diagram is presented in Fig. \ref{block1}.  An external cavity laser is used to generate the optical carrier, which is split into two orthogonal polarizations. Carrier in one of the polarizations is applied to a Thorlabs modulator\,(LN86S-FC) that adds 13\,dB insertion loss. Data signals of amplitude 350\,mV p-p (NRZ in case of QPSK, and PAM-4 in case of 16\,QAM) are applied after pulse shaping to the modulator from a Keysight arbitrary waveform generator (AWG M8195A 65\,GSa/s), which is followed by electrical amplifiers. The modulator output and the carrier are combined using a PBC and sent through a standard 10\,km single mode fiber. 
\par At the receiver end, an EPC (from OZ optics) is used before the PBS. One of the PBS outputs is directly applied to the LO\,(local oscillator) port of the Finisar coherent receiver front-end (CPRV1222A). From the other port, 90\% of the power is applied to the signal port of the coherent receiver front-end and 10\% power is applied to a low bandwidth photo-detector. The photo-detector output is digitized using a data acquisition unit and fed to a controller. The controller changes voltages applied to the EPC according to the gradient descent algorithm that aims at minimizing the photo detector output. This loop continuously adapts the control voltages in real-time to compensate for random fluctuations in the state of polarization. Signals received from the coherent receiver front end are captured using a Keysight real-time oscilloscope.

% needed in second column of first page if using \IEEEpubid
%\IEEEpubidadjcol
\begin{figure}[h!]
	%\vspace{-0.2cm}
	\centering{
		\begin{tabular}{cc}
			\hspace{-0.4cm}\includegraphics[width=4cm,height=3.5cm]{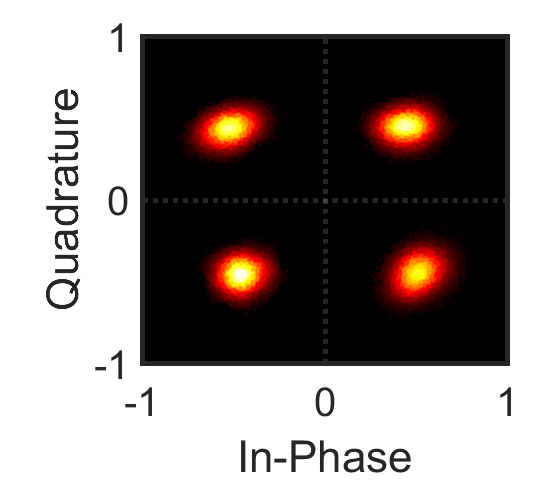}&	\hspace{-0.4cm} \includegraphics[width=4cm,height=3.5cm]{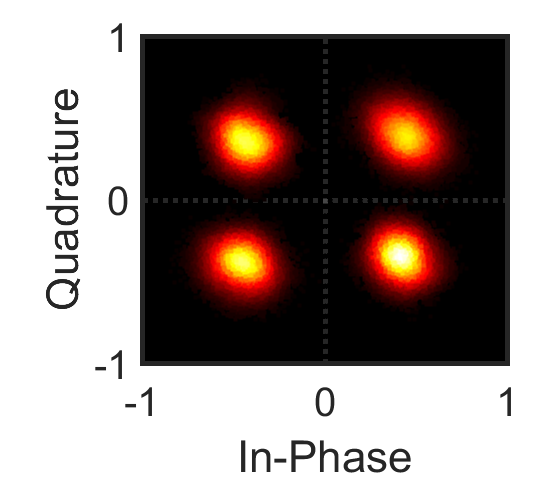} \\
		%	\vspace{-0.2cm}
			\hspace{-0.4cm}(a) &\hspace{-0.4cm} (b) 
	\end{tabular}}
	\caption{ Constellation diagrams for PMC-SH\,QPSK system with 10\,km fiber. (a) Received signal for 10\,Gbaud PMC-SH\,QPSK system having BER: 8.2 $\times$ 10$^{-9}$ without any signal processing and forward error correction\,(FEC); and (b) received signal for 16\,Gbaud PMC-SH\,QPSK system having BER: 5.9 $\times$ 10$^{-5}$ without any signal processing and FEC. }
	\label{r1}
	%\vspace{-0.7cm}		
\end{figure}

\noindent Figure\,\ref{r1}  shows constellation diagrams for 20\,Gbps and 32\,Gbps PMC-SH\,QPSK systems and Fig.\,\ref{r2}  shows constellation diagrams for 40\,Gbps and 64\,Gbps PMC-SH\,16\,QAM systems. Results have been shown without and with offline signal processing. It is evident that no carrier frequency offsets are present (because of which the constellation is stationary even without any signal processing, which mainly involves equalization). Adaptive PC achieves similar or better performance than that achieved using manual control and maintains a constant power  difference of approximately 13\,dB between the two PBS outputs, when the state of polarization is changing randomly at the PBS input. The Receiver front end with monitoring photo detectors are available that can simplify the system further as the monitoring photo detector output can be directly used for feedback. 
\begin{figure}[h!]
	%\vspace{-0.3cm}
	\centering{
		\begin{tabular}{cc}
			\hspace{-0.4cm}\includegraphics[width=4cm,height=3.5cm]{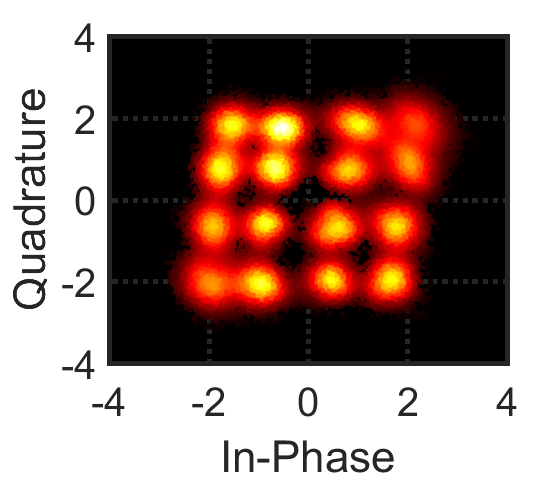}&	\hspace{-0.4cm} \includegraphics[width=4cm,height=3.5cm]{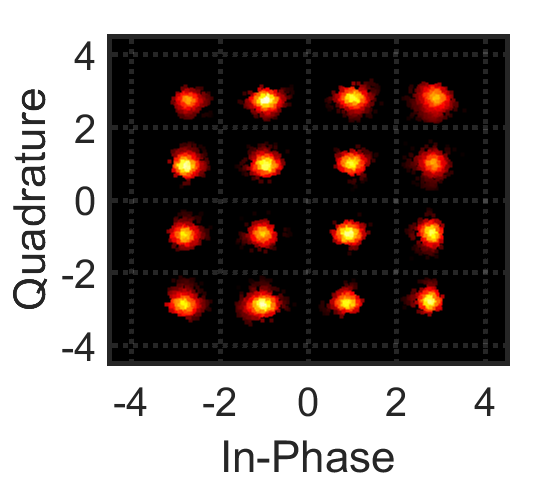} \\
			%\vspace{-0.2cm}
			\hspace{-0.4cm}(a) &\hspace{-0.4cm} (b)
		\end{tabular}}
		\caption{ Constellation diagrams for 10\,Gbaud PMC-SH\,16\,QAM system with 10\,km fiber. (a) Received signal and (b) RDE-DFE equalized signal for 10\,Gbaud PMC-SH\,16\,QAM system having BER: 4.5 $\times$ 10$^{-4}$. RDE: radius directed equalizer and DFE: decision feedback equalizer. }
		\label{r2}
		\vspace{-0.2cm}		
	\end{figure}
	
	\begin{figure}[h!]
		%\vspace{-0.3cm}
		\centering{
			\begin{tabular}{cc}
				%\hspace{-0.4cm}
		      \includegraphics[width=4cm,height=3.5cm]{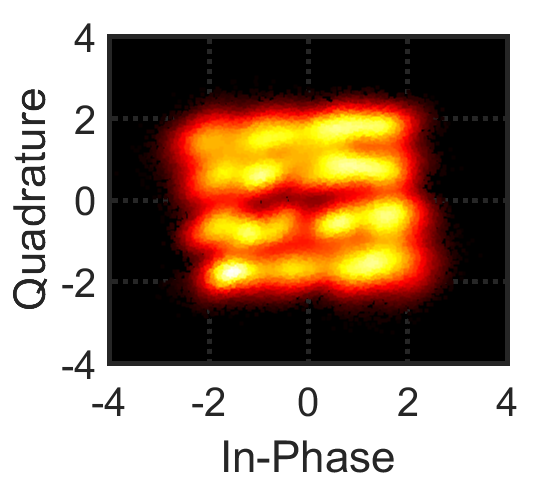}&\hspace{-0.4cm}
				\includegraphics[width=4cm,height=3.5cm]{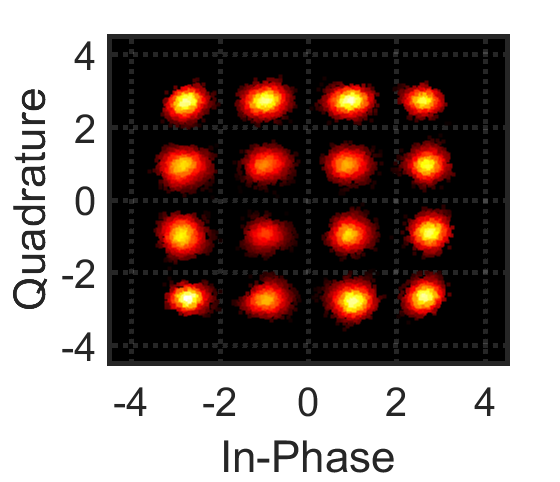} \\
				%\vspace{-0.2cm}
				\hspace{-0.4cm}(a) &\hspace{-0.4cm} (b) 
			\end{tabular}}
			\caption{ Constellation diagrams for 16\,Gbaud PMC-SH\,16\,QAM system with 10\,km fiber: (a) received signal for 16\,Gbaud PMC-SH\,16\,QAM system and (b) RDE-DFE equalized signal for 16\,Gbaud SH-16\,QAM system having BER: 8.7$\times$ 10$^{-3}$. RDE: radius directed equalizer and DFE: decision feedback equalizer. }
			\label{r2}
			%\vspace{-0.2cm}		
		\end{figure}
\section{Conclusion}
In summary, in this work we have presented an adaptive polarization control technique and demonstrated QPSK and 16\,QAM SH systems based on it. This approach makes the SH systems practical and attractive to use for future high-speed data center interconnects.

% if have a single appendix:
%\appendix[Proof of the Zonklar Equations]
% or
%\appendix  % for no appendix heading
% do not use \section anymore after \appendix, only \section*
% is possibly needed

% use appendices with more than one appendix
% then use \section to start each appendix
% you must declare a \section before using any
% \subsection or using \label (\appendices by itself
% starts a section numbered zero.)
%

\section*{Acknowledgment}

The authors would like to thank Meity for funding the project. We would like to thank Dr. Arvind Mishra from Sterlite Technologies for his support.

% Can use something like this to put references on a page
% by themselves when using endfloat and the captionsoff option.
\ifCLASSOPTIONcaptionsoff
  \newpage
\fi

% trigger a \newpage just before the given reference
% number - used to balance the columns on the last page
% adjust value as needed - may need to be readjusted if
% the document is modified later
%\IEEEtriggeratref{8}
% The "triggered" command can be changed if desired:
%\IEEEtriggercmd{\enlargethispage{-5in}}

% references section

% can use a bibliography generated by BibTeX as a .bbl file
% BibTeX documentation can be easily obtained at:
% http://mirror.ctan.org/biblio/bibtex/contrib/doc/
% The IEEEtran BibTeX style support page is at:
% http://www.michaelshell.org/tex/ieeetran/bibtex/
%\bibliographystyle{IEEEtran}
% argument is your BibTeX string definitions and bibliography database(s)
%\bibliography{IEEEabrv,../bib/paper}
%
% <OR> manually copy in the resultant .bbl file
% set second argument of \begin to the number of references
% (used to reserve space for the reference number labels box)
\bibliographystyle{IEEEtran}
\bibliography{references}

\end{document}